\documentclass[a4paper,11pt]{article}
\usepackage{pos}

\title{Expected Constraints on the Intergalactic Magnetic Field using Gamma-Ray Bursts with the Cherenkov Telescope Array Observatory}
\ShortTitle{Expected constraints on the IGMF using gamma-ray bursts with CTAO}

\author*[a]{Ténéman Keita}
\author[b]{Renaud Belmont}
\author[a]{Thierry Stolarczyk}
\affiliation[a]{Université Paris-Saclay, Université Paris Cité, CEA, CNRS, AIM, 91191, Gif-sur-Yvette, France}

\affiliation[b]{Université Paris Cité, Université Paris-Saclay, CEA, CNRS, AIM, 91191, Gif-sur-Yvette, France}

\emailAdd{teneman.keita@cea.fr}
\emailAdd{renaud.belmont@cea.fr}
\emailAdd{thierry.stolarczyk@cea.fr}

\abstract{The InterGalactic Magnetic Field (IGMF), which could permeate the cosmic voids but was never detected so far, is considered a relic of the early Universe. Constraints on its strength $B$ can be derived from its influence on time-delayed very-high-energy photons from Gamma-Ray Bursts (GRBs) in the electromagnetic cascades along their path to the Earth. The present lower limit achieved on its intensity is $10^{-18}\;\mathrm{G}$. In this work, we simulate data from the Cherenkov Telescope Array Observatory (CTAO), accounting for realistic observational constraints, and we apply a joint spectral and temporal fit to characterise the IGMF. GRBs 190114C and 221009A are used as test cases to assess the sensitivity of CTAO. They demonstrate that a broad range of IGMF strengths can be probed with a lower bound as high as $10^{-15}\;\mathrm{G}$. Notably, we show that observations by the CTAO first large telescope, LST-1, already allow us to exclude field strengths up to $3\times 10^{-17}\;\mathrm{G}$.}

\FullConference{
}

\begin{document}
\maketitle

\section{Introduction}

Magnetic fields are observed across all cosmic scales. In most scenarios, their origin requires a pre-existing primordial field amplified through the dynamo effect and/or converging flows of ionized matter. This seed finds its origin in theories related to cosmological phase transitions \cite{Joyce_1997} or inflation \cite{cecchini2023inflationaryhelicalmagneticfields}. The hypothetical magnetic field permeating the cosmic voids is the so-called InterGalactic Magnetic Field (IGMF). It is expected to be the relic of the seed primordial magnetic field, which would have conserved its characteristics in the cosmic voids, far from the influence of galaxy clusters. This IGMF has so far escaped detection. Different mechanisms predict different IGMF properties today, making it a valuable probe of the early Universe.

The IGMF is generally modelled as a turbulent field characterized by a mean strength $B$ in the $10^{-18}$ to $10^{-9}\;\mathrm{G}$ range, and a correlation length $\lambda_B$ spanning $1\;\mathrm{pc}$ to $1\;\mathrm{Gpc}$ \cite{Durrer_2013} (the correlation length is the cosmological distance on which the field intensity and direction can be considered constant). A strong magnetic field would interact with, and induce perturbations in the cosmic plasma detectable in the Cosmic Microwave Background (CMB) power spectrum. It would also turn E-polarized CMB photons into B-polarized photons. Current upper limits on the IGMF intensity are based on the non-observation of those effects.

The IGMF has potentially an observable effect on the very-high-energy (VHE, $\gtrsim 100\;\mathrm{GeV}$) astrophysical source energy spectra and light curves. The photons at $\mathrm{TeV}$ energies from such sources interact with the Extragalactic Background Light (EBL), producing electron-positron pairs. This absorption produces a well-measured exponential cutoff in the flux above around a hundred $\mathrm{GeV}$, depending on the source redshift. These highly relativistic leptons up-scatter CMB photons up to the gamma-ray domain, thus generating a delayed secondary emission at lower energies with respect to the primary photons, and at some angular distance from the source. The absence of detection of this secondary emission can be interpreted as a high-intensity IGMF inducing a large deflection of the leptons, which in turn give gamma-rays that are unobservable on Earth. Lower bounds on $B$ are derived mainly from the absence of expected electromagnetic cascades from extragalactic $\mathrm{TeV}$ sources, like Active Galactic Nuclei (AGNs) and Gamma-Ray Bursts (GRBs).

Because they are persistent sources, AGNs are suited for angular extension studies at the price of the strong assumption that their emission started millions of years ago and no plasma instabilities absorb the initial VHE flux from the source \cite{alawashra2024}. Current AGN-based constraints vary from $10^{-17}\;\mathrm{G}$ to $10^{-12.5}\;\mathrm{G}$ for a $\lambda_B$ value of $1\;\mathrm{Mpc}$.

In contrast, transient sources like GRBs are suited for time-delay searches. Indeed, if the source emission model is known, it is possible to identify the lower energy-delayed secondary photons with respect to the unabsorbed ones. Moreover, their transient nature alleviates the need for assumptions about plasma instabilities or long emission timescales, as the emission duration is measurable and instabilities develop over centuries \cite{alawashra2024}. Recently, GRBs were detected above a few hundred $\mathrm{GeV}$ by Imaging Atmospheric Cherenkov Telescopes (IACTs): H.E.S.S. (GRBs 180720B \cite{Abdalla_2019}, 190829A \cite{Hess_2021} and 221009A \cite{Lhaaso_2023}), MAGIC (GRBs 190114C \cite{2019_bis} and 201216C \cite{Abe_2023}), and more recently the LHAASO wide-field array has detected GRB 221009A \cite{Lhaaso_2023}. First studies on GRBs 190114C and 221009A have given lower limits on $B$ between $10^{-19}$ and $10^{-17}\;\mathrm{G}$ (\cite{Dzhatdoev_2020}, \cite{Dzhatdoev_2023}, \cite{Xia_2024}) using data from MAGIC and LHAASO respectively, as well as unaffected low-energy data from Fermi-LAT allowing for the primary flux characterisation.

The Cherenkov Telescope Array Observatory (CTAO) is the next major advance in VHE astrophysics \cite{hofmann2023cherenkovtelescopearray}. Its combination of Large, Medium, and Small Sized Telescopes allows an energy span from $20\;\mathrm{GeV}$ to $300\;\mathrm{TeV}$. With superior sensitivity near $100\;\mathrm{GeV}$, CTAO has the potential to improve IGMF constraints. A first publication based on an idealised detection of GRBs 190114C and 221009A indicates that CTAO could challenge the best limits obtained from time-delayed emission \cite{Miceli_2024}. However, the influence of realistic observational conditions like the conditions of dark time, the reduction of the telescope efficiency at low altitudes, and the increase of the effective energy threshold have never been considered. This is the objective of the present work.

\section{Method}

Our work is based on an electromagnetic cascade simulation code affecting the initial GRB flux, the CTAO response functions and a likelihood model fit to characterize the presence of a delayed low-energy photon excess.

\subsection{Theory}

\begin{figure}
    \centering
    \includegraphics[width=0.75\linewidth]{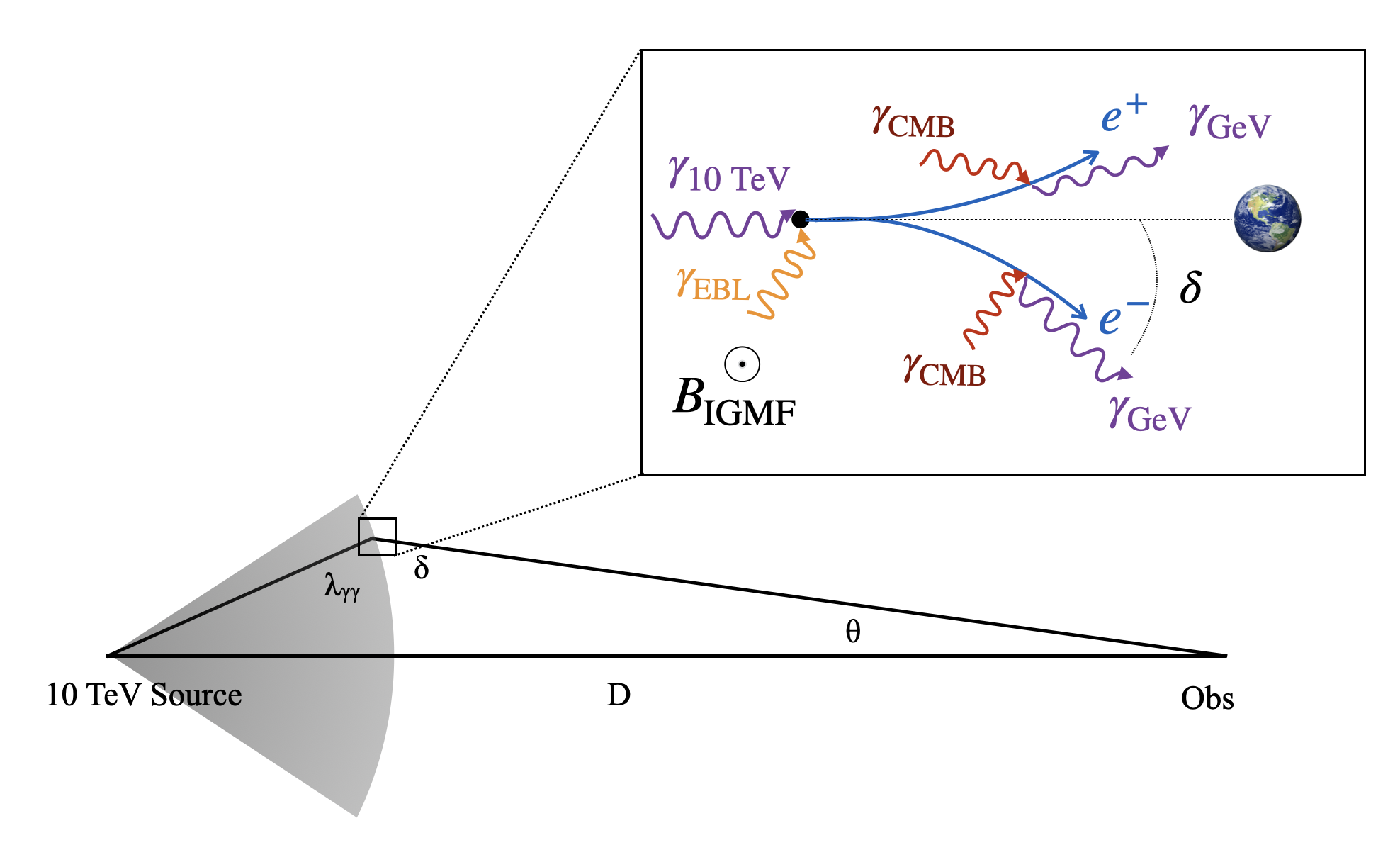}
    \caption{Schematic representation of the cascade process in the one-generation approximation: primary photons generate electron-positron pairs at a distance $\lambda_{\gamma\gamma}$ from the source. These pairs are deflected by the IGMF by an angle $\delta$, and subsequently emit secondary photons that are observed at an angle $\theta$ relative to the line of sight.}
    \label{triangle}
\end{figure}

The cosmological cascade mechanism is schematized in Fig.\,\ref{triangle}. After travelling $\lambda_{\gamma\gamma} \sim 100\;\mathrm{Mpc}$, $\mathrm{TeV}$ photons interact with EBL photons and produce relativistic lepton pairs. The IGMF induces a deflection $\delta$ along the path of the lepton. Assuming a source distance $D \sim 1\;\mathrm{Gpc} \gg \lambda_{\gamma\gamma}$, and a small angle $\delta$, the time delay $\Delta t$ relates to $\delta$ by $c\Delta t \approx \lambda_{\gamma\gamma}\delta^2/2$ \cite{Fitoussi2017bis}. The secondary photons produced by Compton scattering depend on the lepton energy squared. Using the dependency of $\delta$ to the IGMF for large correlation lengths, and the lepton energy, we can express the typical secondary photon energy $\langle E_\gamma \rangle$ as:
\begin{equation}
    \langle E_\gamma \rangle = \left(\frac{B}{10^{-17}\;\mathrm{G}}\right) \left(\frac{\lambda_{\gamma\gamma}}{100\;\mathrm{Mpc}}\right)^{1/2} \left(\frac{\Delta t}{2\;\mathrm{days}}\right)^{-1/2} \mathrm{TeV}
    \label{energy}
\end{equation}
The mean secondary gamma-ray energy is higher when the magnetic field is higher. The Fermi-LAT sensitivity drops above $\sim10\;\mathrm{GeV}$, ideal for constraining $B \leq 10^{-19}\;\mathrm{G}$ \cite{Dzhatdoev_2020}. IACTs detect energies up to the $\mathrm{TeV}$, but over shorter timescales. As a consequence, CTAO and other IACTS can probe stronger fields depending on their sensitivity.

\subsection{Simulation code}

We use the full 3D in space Monte Carlo code \verb|CascadEl| \cite{Fitoussi2017bis}, which simulates photon and lepton propagation from a distant $\mathrm{TeV}$ source to Earth for a given $(B,\;\lambda_B,\;z)$. The simulation takes into account a $\Lambda\mathrm{CDM}$ cosmology $(H_0=68.7, \;\Omega_m=0.3,\; \Omega_\lambda=0.7)$. It models photon-photon pair production and Compton scattering on the CMB and EBL \cite{dominguez2011} whose evolution with redshift is taken into account. The interactions are generated according to the exact first-order quantum electrodynamics cross sections, including Klein-Nishina for leptons, with the appropriate sampling of the energies and the directions. The leptons follow helical paths in a cubic-cell IGMF model with uniform field strength but random orientation per cell of size $\lambda_B$. Results are averaged over $100$ realisations of the magnetic field orientations. The simulation tracks particles until they cross an observer sphere at a given comoving distance to the source corresponding to its redshift. At this point, the energy, the time delay, the angular position, the nature of the particle (photon or lepton), the rank of interactions in the cascade for that particle, as well as the primary photon energy are stored for the next steps. Each photon will then have a weight based on the energy, the emission time and the emission angle of their primary ancestor photon, to account for any source model at a given redshift and a given IGMF configuration. 

With this code, we generate a photon list at the level of the CTAO arrays assuming a primary photon flux $\Phi\equiv\mathrm{d}N/\mathrm{d}E$ (in $\mathrm{ph/erg/s}$) as a power-law in energy and time with an exponential cutoff energy:
\begin{equation}
    \Phi(E,t>t_\mathrm{min}) = \Phi_0  E^{-\gamma}t^{-\alpha}\exp(-E/E_\mathrm{cut})
\end{equation}
The CTAO response is simulated using the \verb|Gammapy| software \cite{Donath_2023} and the official CTAO Instrument Response Functions (IRFs)\footnote{CTAO Instrument Response Functions - prod5 version v0.1, https://doi.org/10.5281/zenodo.5499840} for the alpha configuration. This configuration comprises $4$ Large size telescopes (LSTs) and $9$ medium size telescopes (MSTs) in the North (La Palma Canaries Island, Spain), and $14$ MSTs and $37$ Small size telescopes (SSTs) in the South (Paranal, Chile). LSTs have been specially designed for catching GRBs with a maximal required slewing time of $30\;\mathrm{s}$ and an energy threshold of $20\;\mathrm{GeV}$ at high altitudes and $110\;\mathrm{GeV}$ down to $30^\circ$ above the horizon (energy thresholds are obtained for $10\%$ of the peak effective area). MSTs have a maximal required slewing time of $90\;\mathrm{s}$ and an energy threshold varying from $60$ to $200\;\mathrm{GeV}$ from high to low altitudes. The IRFs are defined at fixed zenith angles $(20^\circ, 40^\circ, 60^\circ$) corresponding to observations in optimal and stable conditions (dark nights, no clouds) with a minimal altitude of $24^\circ$. We have ignored the presence of Moonlight. The corresponding data are analysed using a reflected region technique \cite{Berge_2007}. To extract both intrinsic properties and IGMF strength, we use table models and perform a joint likelihood fit of all time bins with five free parameters: $\gamma$, $\alpha$, $E_\mathrm{cut}$ and $B$ and a normalization factor $\Phi_0$. We impose a lower limit of $10\;\mathrm{TeV}$ for the fitted cutoff energy, which is necessary to avoid a degeneracy between a low $E_\mathrm{cut}$ and a strong $B$ in the case of a non-detection of the cascade. Finally, the correlation length is set to $1\;\mathrm{Mpc}$.

\section{Preliminary results}

We have applied the methodology described above to two of the five GRBs detected so far in the $\mathrm{TeV}$ energy range: GRB 190114C and GRB 221009A.

\subsection{GRB 190114C}

\begin{figure}
    \centering
    \includegraphics[width=0.75\linewidth]{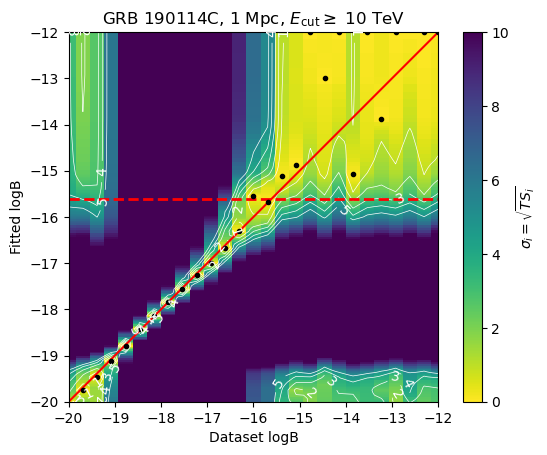}
    \caption{Fit map at $\lambda_B=1\;\mathrm{Mpc}$. The x-axis represents the injected values of $B$, and the corresponding best-fit values are shown as black dots on the y-axis. The colour scale denotes the confidence level associated with each fit. If every fit was totally accurate, the black dots would all lie on the red diagonal $\log B_\mathrm{fit} = \log B_\mathrm{dataset}$.}
    \label{emax}
\end{figure}
GRB 190114C is a bright long GRB at redshift $z=0.425$, initially detected by both Fermi-GBM and Swift-BAT. The Fermi data analysis estimates the isotropic energy at $2.5 \times 10^{53}\;\mathrm{erg}$ in the $[1,\,10^4]\;\mathrm{keV}$ band \cite{2019_bis}. MAGIC observed the GRB in the $0.1$–$10\;\mathrm{GeV}$ range from $T_0 + 62\;\mathrm{s}$ to $\sim T_0 + 2400\;\mathrm{s}$, where $T_0$ is the Fermi-GBM burst time, and extended the observation at a very low altitude. Synthetic data for this GRB are produced assuming:
\begin{equation}
    \Phi(E,t>t_\mathrm{min}) = \Phi_0 \left(\frac{E}{1 \; \mathrm{TeV}}\right)^{-2.22}\left(\frac{t}{100 \; \mathrm{s}}\right)^{-1.60}\;\exp \;(-E/E_\mathrm{cut})
\end{equation}
with $E_\mathrm{cut}=10 \; \mathrm{TeV}$ as a conservative assumption justified by the lack of data at high energy. Here, we follow the assumption of \cite{ravasio_2019} and assume that $t_\mathrm{min}=6\;\mathrm{s}$. $\Phi_0$ is calibrated such that the time-integrated flux matches the MAGIC observation: $\int_{62}^{2400}\Phi(E,t)\;\mathrm{d}t = 1.07\times 10^{-8}\;\mathrm{TeV}^{-1}\;\mathrm{s}^{-1}\;\mathrm{cm}^{-2}$ at $E = 423.8\;\mathrm{GeV}$. At the CTAO North site, the burst appears at night and remains visible for $3$ hours the first night and $4$ hours the subsequent nights, although at low altitudes thus constraining a reconstructed energy threshold of $110\;\mathrm{GeV}$. We impose a $60\;\mathrm{s}$ time delay before starting the observation to account for the alert delay and telescope re-pointing. In the CTAO South site, the observation starts $2$ hours after the alert and lasts $5$ hours at high altitudes, allowing reconstruction down to $60\;\mathrm{GeV}$. In these conditions, this GRB is detected at $1388\sigma$ in the North at $T_0+30\;\mathrm{min}$ and $67\sigma$ in the South at $T_0+9$ hours. It is noteworthy that the maximal significance of the IGMF signature generally occurs later than the peak significance of the GRB itself, typically after the first few hours or nights. We analyse data up to the $5^\mathrm{th}$ night since the IGMF effect is only visible up to the $5^\mathrm{th}$ night for the strongest detectable magnetic fields. We verified that more nights would accumulate more noise than signal, worsening the results. Whereas the results obtained using less than $5$ nights were also worse.

The results of the joint (North and South) fit are summarized in Fig.\,\ref{emax}. This colour map shows the confidence levels for the logarithm of the magnetic intensity for different IGMF discrete values. In the region $10^{-19}\;\mathrm{G} \lesssim B \lesssim 10^{-16}\;\mathrm{G}$, the fit converges to the IGMF value used to produce the data set because the secondary photon excess falls into the CTAO energy range and observation windows. Stronger fields ($B > 10^{-16}\;\mathrm{G}$) dilute the secondary emission so much that it remains undetectable. Based on the $3\sigma$ contours on Fig.\,\ref{emax}, CTAO allows getting lower limits as high as $\sim 10^{-16}\;\mathrm{G}$ in this particular case (red dashed line). These results are quite insensitive to $E_\mathrm{cut}$ because GRB 190114C has a soft spectrum ($\gamma = 2.22$), and the intrinsic flux above $10\;\mathrm{TeV}$ is negligible, regardless of the position of the cutoff.

\subsection{GRB 221009A}

\begin{figure}
    \centering
    \includegraphics[width=0.49\linewidth]{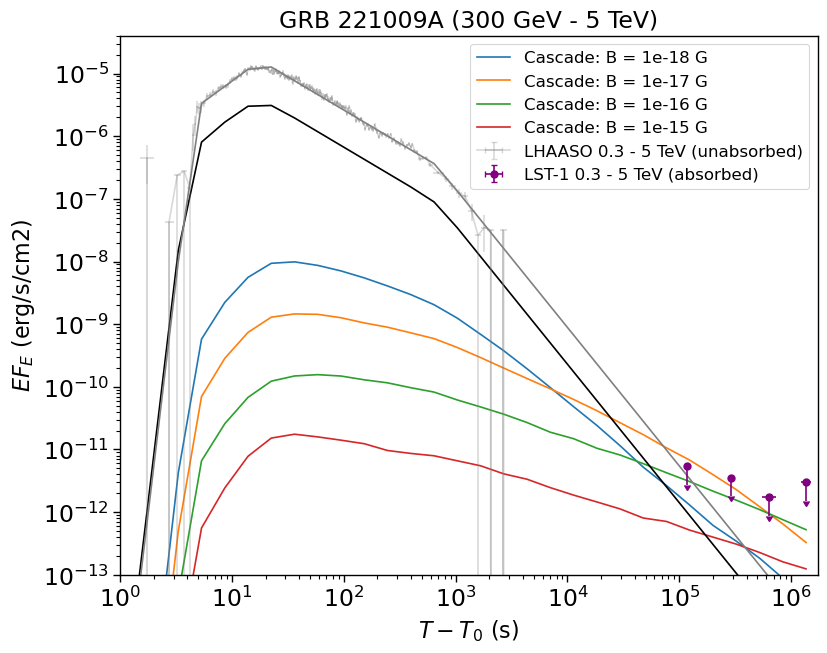}
    \includegraphics[width=0.47\linewidth]{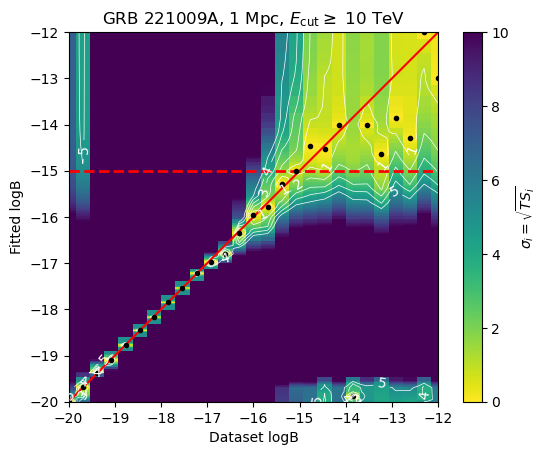}
    \caption{(a) Light curve of GRB 221009A with the primary emission (black line) and the cascade for various $B$ (colours). The LST-1 upper limits are shown in purple. (b) Fit map at $\lambda_B=1\;\mathrm{Mpc}$.}
    \label{lightcurve}
\end{figure}

GRB 221009A was first detected by Fermi-GBM at a redshift of $z=0.151$. It is the brightest GRB to date, with $E_{\mathrm{iso}} \sim 10^{55}\;\mathrm{erg}$ \cite{Lhaaso_2023}. A VHE emission was detected by LHAASO from the burst time to $6000\;\mathrm{s}$. The prototype LST of CTAO, LST-1, was not active during the first night, and later observations between $T_0 + 1.33\;\mathrm{days}$ and $T_0 + 19\;\mathrm{days}$ gave preliminary upper limits \cite{lst1}. Synthetic data for this GRB were produced assuming:
\begin{equation}
    \Phi(E,t) = \Phi_0 \left(\frac{E}{1 \; \mathrm{TeV}}\right)^{-2.32}f_\mathrm{BPL}(t,t_i,\alpha_i)\;\exp \;(-E/E_\mathrm{cut})
\end{equation}
with $\Phi_0$ normalized to match the LHAASO flux at $t = 18.8\;\mathrm{s}$:
$\int_{0.3\;\mathrm{TeV}}^{5\;\mathrm{TeV}} \Phi(E,t)\;\mathrm{d}E = 1.31 \times 10^{-5}\;\mathrm{erg}^{-1}\;\mathrm{s}^{-1}\;\mathrm{cm}^{-2}$, and $f_\mathrm{BPL}(t)$ a broken power-law with five temporal segments (rising and decaying phases), namely: $-14.9,\; -1.30,\; 0.05,\; 1.06,\; 2.21$, the latter being used for longer times. If the full CTAO array would have been active, the GRB would have become visible $7$ hours after the alert at a favourable altitude ($\gtrsim 40^\circ$) in the North site, allowing reconstruction down to $30\;\mathrm{GeV}$. The observation in the South site would have began $11$ hours after $T_0$ and at a lower altitude and a higher energy threshold ($350\;\mathrm{GeV}$). Under these conditions, the detection occurs in CTAO North and South with maximal significances of $65\sigma$ and $64\sigma$ reached $11$ hours and $12$ hours after the burst, respectively. However, regarding the characterization of the IGMF which can occur hours or nights after the peak significance depending on $B$, and due to the lower energy threshold in the South, cumulating data from North and South did not improve the North result. 

GRB 221009A has been analysed up to $12$ nights, beyond which the flux falls below the CTAO North array sensitivity for even the strongest detectable fields.  Fig.\,\ref{lightcurve}\,(a) presents the expected light curve in the available energy range in the Northern site, and for magnetic field strengths in the range from $10^{-18}$ to $\sim 10^{-15}\;\mathrm{G}$. Despite the LST-1 observation starting $1.33\;\mathrm{days}$ after the burst, its upper limits already exclude $B \in [3 \times 10^{-18},\;3 \times 10^{-17}]\;\mathrm{G}$ assuming $E_\mathrm{cut} = 10\;\mathrm{TeV}$ (a stronger cutoff energy would induce even stricter constraints). We reproduced the same analysis as for GRB 190114C with the full CTAO array. From the resulting colour map on Fig.\,\ref{lightcurve}\,(b), we can deduce that CTAO would be capable of placing robust lower limits up to $\sim 10^{-15}\;\mathrm{G}$ even assuming a conservative $E_\mathrm{cut}=10\;\mathrm{TeV}$.

\section{Conclusion}

In the present study, we show that CTAO can improve the current lower bounds on the IGMF even in accounting for realistic observational constraints, such as a limited dark time, source altitude and varying energy thresholds. A GRB like 190114C would allow constraining the magnetic fields up to $10^{-16}\;\mathrm{G}$ for a correlation length of $1\;\mathrm{Mpc}$. Regarding the GRB 221009A, current LST-1 upper limits already exclude magnetic fields between $3\times 10^{-18}$ and $3\times 10^{-17}\;\mathrm{G}$ under the conservative assumption that $E_\mathrm{cut} = 10\;\mathrm{TeV}$ while an observation with the CTAO full array would have reached $10^{-15}\;\mathrm{G}$. Such lower limit would make it possible to exclude most inflationary magnetogenesis scenarios, as well as all chiral models capable of producing the observed baryon asymmetry \cite{neronov2020viscousdampingchiraldynamos}. A direct detection could provide indirect support for such models.

\section*{Acknowledgment}

\noindent We gratefully acknowledge financial support from the agencies and organizations listed at \href{https://www.ctao.org/for-scientists/library/acknowledgments/}{https://www.ctao.org/for-scientists/library/acknowledgments/}. This research has made use of the CTA instrument response functions provided by the CTA Consortium and Observatory, see \href{https://www.cta-observatory.org/science/ctao-performance}{https://www.cta-observatory.org/science/ctao-performance} for more details.

\end{document}